\documentclass[12pt,preprint]{aastex}

\newcommand{\msun}{\thinspace M_\odot}  
\newcommand{\rsun}{\thinspace R_\odot}

\newcommand{\nc}{n_{\rm c}}

\newcommand{\cm  }{\,{\rm cm}^{-3} }

\newcommand{\km  }{\,{\rm km\, s^{-1}} }

\begin{document}

\title{Protostellar Jet and Outflow in the Collapsing Cloud Core}
\author{Masahiro N. Machida, Shu-ichiro Inutsuka, and Tomoaki Matsumoto}

\begin{abstract}
Using three-dimensional resistive MHD nested grid simulations, we investigate the driving mechanism of outflows and  jets in star formation process.
Starting with a Bonnor-Ebert isothermal cloud rotating in a uniform magnetic field, we calculated cloud evolution from the molecular cloud core ($\nc=10^4 \cm$, $r=4.6\times 10^4$\,AU) to the stellar core ($\nc=10^{22} \cm$, $r \sim 1 R_\odot$), where $\nc$ and $r$ denote the central density and radius of each object, respectively. 
In the collapsing cloud core, we found two distinct flows:
Low-velocity outflows ($\sim$5$\km$) with a wide opening angle, driven from the adiabatic core, and high-velocity jets ($\sim$30$\km$) with good collimation, driven from the protostar.
High-velocity jets are enclosed by low-velocity outflow.
The difference in the degree of collimation between the two flows is caused by the strength of the magnetic field and configuration of the magnetic field lines.
The magnetic field around an adiabatic core is strong and has an hourglass configuration; therefore, the low-velocity outflow from the adiabatic core are driven mainly by the magnetocentrifugal mechanism and guided by the hourglass-like field lines.
In contrast, the magnetic field around the protostar is weak and has a straight configuration owing to Ohmic dissipation in the high-density gas region.
Therefore, high-velocity jet from the protostar are driven mainly by the magnetic pressure gradient force and guided by straight field lines. 
Differing depth of the gravitational potential between the adiabatic core and the protostar cause the difference of the flow speed.
Low-velocity outflows correspond to the observed molecular outflows, while high-velocity jets correspond to the observed optical jets.
We suggest that the protostellar outflow and the jet are driven by different cores, rather than that the outflow being entrained by the jet. 
\end{abstract}

\section{Introduction}
\label{sec:1}
The observations indicate that outflow is ubiquitous in the star formation process, and flows from the protostars have two or more distinct velocity components \cite{hirth97,pyo03}.
Typically, a flow from the protostar is composed of a low-velocity component (LVC) with 10-50$\km$ and a high-velocity component (HVC) with $\sim$$100\km$.
There is a clear trend toward higher collimation at higher flow  velocity \cite{arce06}. 
Since flows from the protostar have various morphological and kinematical properties, they cannot be explained by a single-class model.
The flows that originated from the protostar are typically classified into two types: molecular outflow observed mainly with CO molecules \cite{arce06}, and optical jet observed by optical emission \cite{pudritz06}.
Molecular outflows observed by CO line emission exhibit a wide opening angle \cite{belloche02} and slower velocity of $10-50\km$ \cite{arce06}, while optical jets observed by optical emission exhibit good collimation  and higher velocity  of $100-500\km$ \cite{bally07}.
Observations indicate that around each protostar, high-speed jets with a narrow opening angle are enclosed by a low-velocity outflow with a wide opening angle \cite{mundt83}. 
However, the driving mechanism of these flows are still unknown.

In this study, we calculate cloud evolution from the molecular cloud core ($n_c = 10^4\cm$, $r_c = 4.6 \times 10^4$\,AU) to stellar core formation ($n_c \simeq 10^{22} \cm$, $r_c \simeq 1 \rsun$) using three-dimensional resistive MHD nested grid method,  study the formation process of  jets and  outflows, and show the driving mechanisms of these flows.

\section{Model}
 Our initial settings are almost the same as those of \cite{machida06a,machida06b,machida07,machida07c,machida08}.
We solve the resistive MHD equations including self-gravity (see, eqs.1-5 of \cite{machida07}).  
We adopt a spherical cloud with critical Bonnor-Ebert  density profile having $\rho_{\rm BE} = 3.841 \times 10^{-20} \, \rm{g} \, \cm$ ($n_{\rm BE} = 10^4\cm$) of the central (number) density as the initial condition.
 The critical radius for a Bonnor--Ebert sphere $R_c = 6.45\, c_s [4\pi G \rho_{\rm {BE},0}]^{-1/2}$ corresponds to $ R_c = 4.58 \times 10^4$\,AU for our settings. 
 Initially, the cloud rotates rigidly ($\Omega_0=7\times 10^{15}$\,s$^{-1}$) around the $z$-axis and has a uniform magnetic field ($B_0 =17 $$\mu$G) parallel to the $z$-axis (or rotation axis).
To promote contraction, we increase the density by 70\% from the critical Bonnor-Ebert sphere.
The initial central density is therefore $\rho_0 = 6.53 \times 10^{-20}$\,g$\cm$ ($n_0 = 1.7\times 10^{-4}\cm$).

We adopt the nested grid method  \cite{machida06a,machida04,machida05a} to obtain high spatial resolution near the center.
Each level of a rectangular grid has the same number of cells ($ 64 \times 64 \times 32 $),  although the cell width $h(l)$ depends on the grid level $l$.
The highest level of a grid changes dynamically.
The box size of the initial finest grid $l=1$ is chosen to be $2^4 R_{\rm c}$, where $R_c$  denotes the radius of the critical Bonnor--Ebert sphere. 
 A new finer grid is generated whenever the minimum local  Jeans length $ \lambda _{\rm J} $ becomes smaller than $ 8\, h (l_{\rm max}) $. 
The maximum level of grids is restricted to $l_{\rm max} = 30$.

\begin{figure}
\includegraphics[width=140mm]{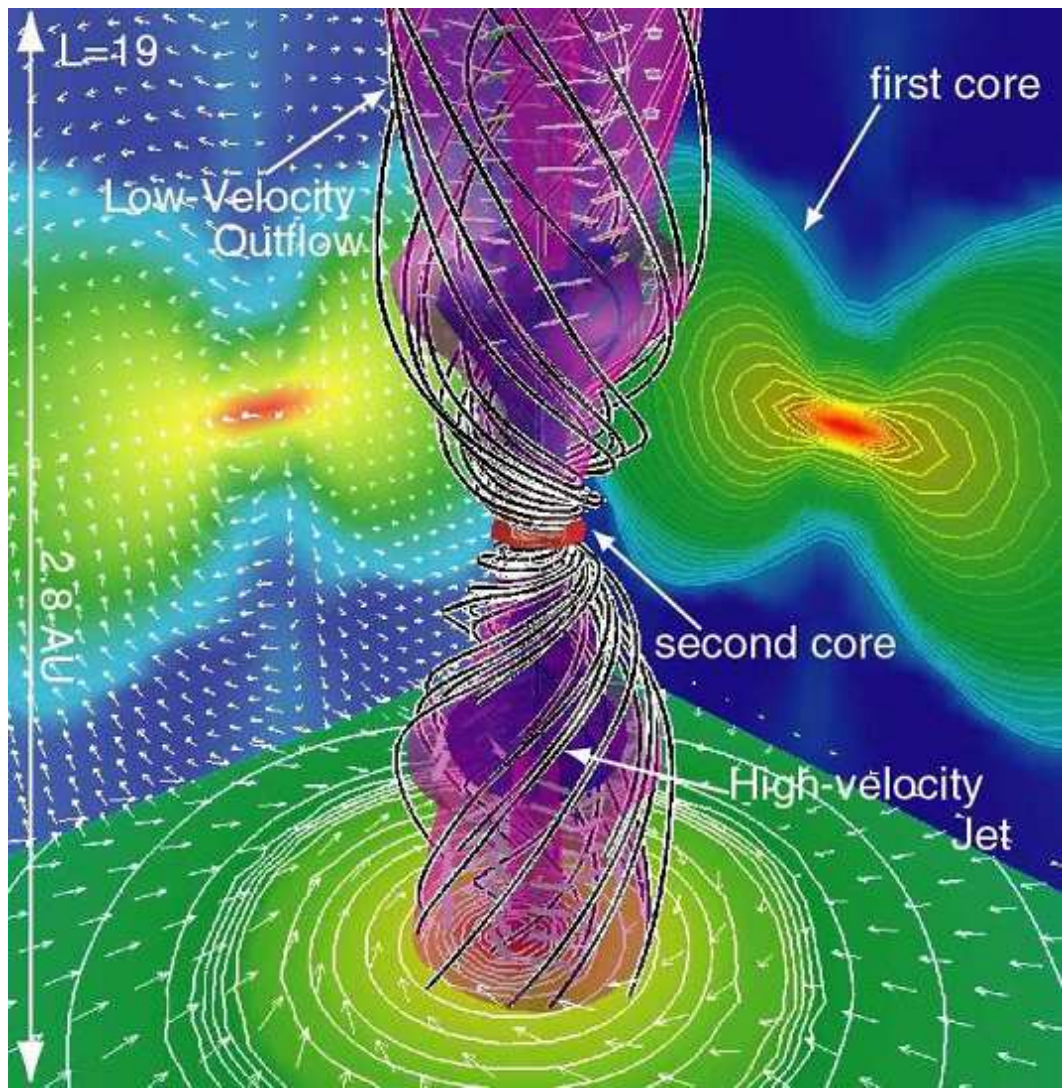}
\caption{
Bird's-eye view.
The structure of high-density region ($\rho > 0.1\rho_{\rm c}$; red iso-density surface),  and magnetic field lines (black-and-white streamlines) are plotted in each panel.
The structures of the jet ($v > 7\km$) and outflow ($v > 0.5\km$) are shown by iso-velocity surfaces, respectively.
The density contours (false color and contour lines), velocity vectors (thin arrows) on the mid-plane of $x=0$, $y=0$, and $z=$0 are, respectively, projected in each wall surface.
}
\label{fig:1}       
\end{figure}

\section{Results}
The molecular gas obeys the isothermal equation of state with temperature of $\sim 10$ K until $n_c \simeq 5 \times 10^{10}\cm$ (isothermal phase), then cloud collapses almost adiabatically  ($5\times 10^{10}\cm < n_c < 10^{16}\cm$; adiabatic phase) and quasi-static core (i.e., first core)  forms during the adiabatic phase \cite{larson69, masunaga98}.
In our calculations,  the first core forms when the central density reaches $n_c \simeq 8\times10^{12} \cm$.
The magnetic flux is removed from the  first core during the adiabatic phase by the Ohmic dissipation \cite{nakano02}.
After central density reaches $n_c \simeq 10^{16}\cm$, the equation of state becomes soft reflecting the dissociation of hydrogen molecules at $T \simeq 2\times 10^3$ K, and collapses rapidly.
By this epoch, the central temperature becomes so high that the thermal ionization of Alkali metals reduces the resistivity and so that Ohmic dissipation becomes ineffective.
Thus, the magnetic field becomes strong again as central region collapses.
The second core (or protostar) \cite{larson69} forms at $n_c \simeq 10^{21} \cm$.
The magnetic field strength increases rapidly after the second core formation epoch ($n >10^{21} \cm$), because the shearing motion between the second core and ambient medium amplifies the toroidal magnetic field around the second core.

Figure~\ref{fig:1} shows the structure of the low- and high-velocity flows, and the configuration of the magnetic field lines.
It also shows the shapes of the first core (left panel; the projected density contours on the wall) and the protostar (right panel; the red isosurface).
The purple and blue surfaces in Figure~\ref{fig:1} indicate the iso-velocity surface of $v_z = 5 \km$ and $v_z = 0.5 \km$, respectively.
The flow inside the purple iso-velocity surface has a velocity of $v > 0.5\km$ (low-velocity component; LVC),  while the flow inside the blue iso-velocity surface has a velocity of $v>5\km$ (high-velocity component; HVC).
The HVC is enclosed by the LVC.
The LVC flow is mainly driven from the first core, while the HVC flow is mainly driven from the protostar.
The LVC and HVC are strongly coiled by the magnetic field lines anchored to the first core and protostar, respectively.

\begin{figure}
\includegraphics[width=150mm]{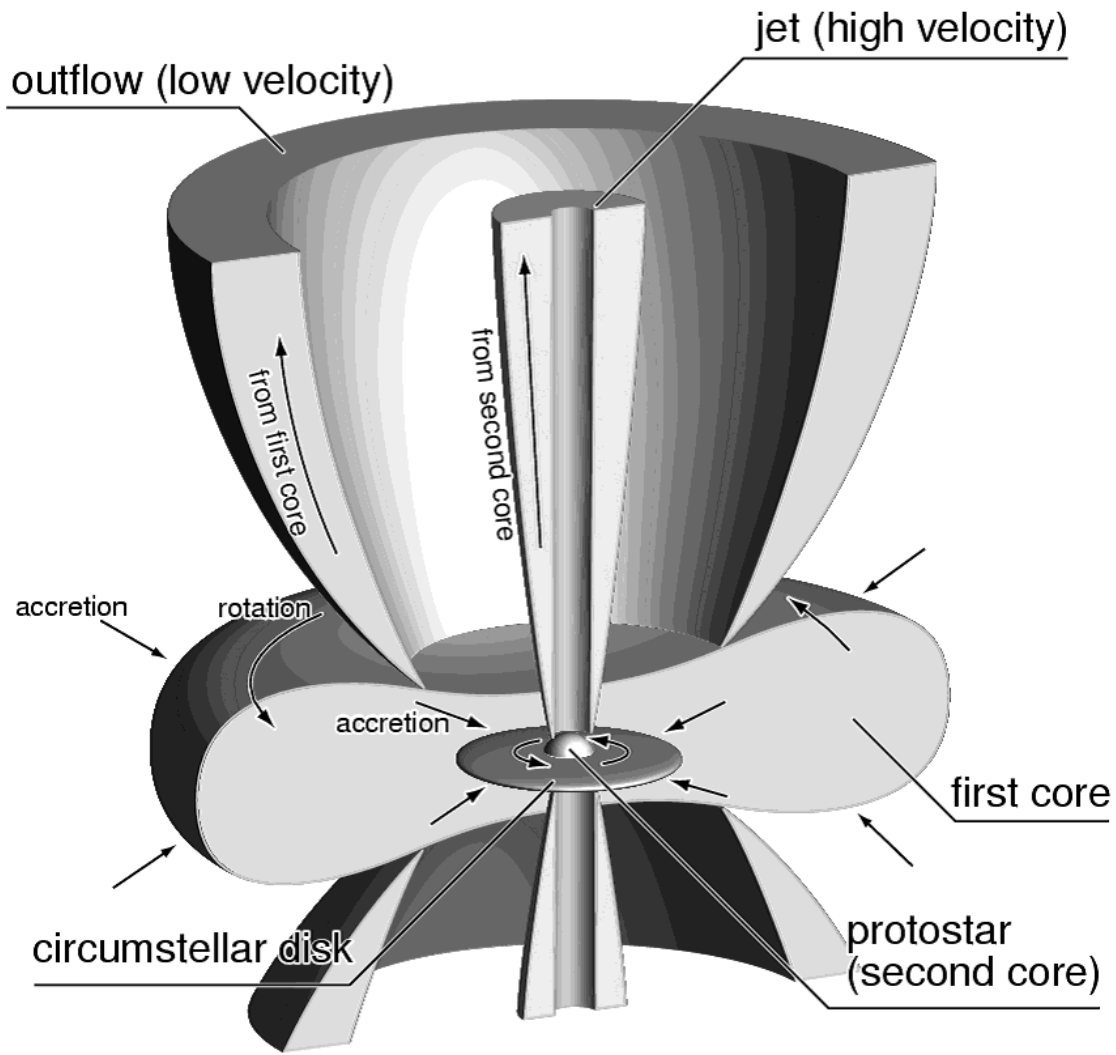}
\caption{
Schematic view of the jet and outflow driven from the protostar and the first core, respectively.
}
\label{fig:2}       
\end{figure}

Our results show that the flow appearing around the first core has a wide opening angle and slow speed, while the flow appearing around the protostar has a well-collimated structure and high speed,  as shown in Figure~\ref{fig:2}.
The speed difference is caused by the difference of the depth in the gravitational potential.
The flow speed corresponds to the Kepler speed of each object.
Because the first core has a shallow gravitational potential, its flow is slower.
The flow driven from the protostar, which has a deeper gravitational potential, has a high speed.
In our calculations, the low- and high-velocity flows have speeds of $v_{\rm LVF} \simeq 3\km$ and $v_{\rm HVF}\simeq 30\km$, respectively.
These speeds are slower than those of observations. 
Typically, observed molecular outflow and optical jet have speeds of  $v_{\rm out,obs} \simeq 30\km$, and $v_{\rm jet,obs}\simeq 100\km$, respectively.
However, since the first and second cores (protostar) have mass of $M_{\rm fc} = 0.01\msun$ and $M_{\rm sc} \simeq 10^{-3}\msun$, respectively, at the end of the calculations, each core increases its mass in the gas accretion phase. 
The Kepler speed increases with the square root of the mass.
When the mass of each core increases by 100 times, the Kepler speed increases 10 times.
Thus, the speed of the low- and high-velocity flows  may increase by 10 times, and reach  $v_{\rm LVF} \simeq 30\km$ and $v_{\rm HVF}=300\km$, respectively, which correspond to typical observed values.

\section{Discussion}
Observation shows that the molecular outflows have wide opening angles and low flow speeds, while the optical jets have good collimation and high flow speeds.
Molecular outflow has been considered to be entrained by the optical jet driven from a circumstellar disk around the protostar.
In this study, we calculated the cloud evolution from the molecular cloud core to protostar formation, and found that two distinct flows are driven from different objects, and the observed features of molecular outflow and optical jet  were naturally reproduced.
Thus, we expect that the low-velocity flow  from the first core corresponds to the molecular outflow, while the high-velocity flow from the protostar corresponds to the optical jet.

The different collimation of  low- and high-velocity flow is caused both by the configuration of the magnetic field lines around the drivers and their driving mechanisms.
The magnetic field lines around the first core have an hourglass configuration because they converge to the cloud center as the cloud collapses, and Ohmic dissipation is ineffective before the first core formation.
In addition, the centrifugal force is more dominant than the Lorentz force in the low-velocity flow (molecular outflow). 
Thus, the flow appearing near the first core is mainly driven by the magnetocentrifugal wind mechanism (disk wind).
On the other hand, near the protostar, the magnetic field lines are a straight, and the magnetic pressure gradient mechanism is more effective for driving the high-velocity flow (optical jet).
The magnetic field lines straighten by the magnetic tension force near the protostar because the magnetic field is decoupled from the neutral gas.
However, the magnetic field lines are strongly twisted in the region in close proximity to the protostar, where the magnetic field is coupled with the neutral gas again.
Thus, the strong toroidal field generated around the protostar can drive the high-velocity flow (optical jet), which is guided by the straight configuration of the magnetic field.

Our calculations do not completely reject the well-known concept that the observed molecular outflow is entrained by the optical jet, because we calculate the formation of the  jet and  outflow only in the early star-formation phase.
Further long-term calculations are needed to understand the mechanism of the optical jet and molecular outflow in more detail.

\end{document}